\documentclass[traditabstract,referee]{aa}

\usepackage{txfonts}
\usepackage{natbib}
\bibpunct{(}{)}{;}{a}{}{,}
\usepackage{graphicx}

\begin{document}

\title{Large-scale
  3D MHD
  simulation
  on the solar flux emergence
  and the small-scale dynamic features
  in an active region}
\author{S.~Toriumi \and T.~Yokoyama}
\institute{Department of Earth and Planetary Science,
  University of Tokyo,
  7-3-1 Hongo, Bunkyo-ku, Tokyo 113-0033, Japan\\
  \email{toriumi@eps.s.u-tokyo.ac.jp}
}
\date{Received / Accepted}

\abstract{
  We have performed
  a three-dimensional
  magnetohydrodynamic simulation
  to study the emergence
  of a twisted magnetic flux tube
  from $-20,000\ {\rm km}$
  of the solar convection zone
  to the corona
  through the photosphere
  and the chromosphere.
  The middle part
  of the initial tube
  is endowed with a density deficit
  to instigate a buoyant emergence.
  As the tube approaches the surface,
  it extends horizontally
  and makes a flat magnetic structure
  due to the photosphere
  ahead of the tube.
  Further emergence
  to the corona
  breaks out
  via the interchange-mode instability
  of the photospheric fields,
  and eventually several magnetic domes
  build up above the surface.
  What is new
  in this three-dimensional
  experiment is,
  multiple separation events
  of the vertical magnetic elements
  are observed
  in the photospheric magnetogram,
  and they reflect
  the interchange instability.
  Separated elements
  are found to
  gather at the edges
  of the active region.
  These gathered elements
  then
  show shearing motions.
  These characteristics
  are highly reminiscent
  of active region observations.
  On the basis
  of the simulation results above,
  we propose
  a theoretical picture
  of the flux emergence
  and the formation of
  an active region
  that explains the observational features,
  such as multiple separations of faculae
  and the shearing motion.
}


\authorrunning{S.~Toriumi \& T.~Yokoyama}
\titlerunning{3D MHD simulation on the solar flux emergence}

\maketitle

\section{Introduction\label{sec:intro}}

Solar active regions including sunspots
are generally thought to be
the consequence of the magnetic flux emergence
\citep{par55}.
Observationally,
they appear 
as a developing bipolar pair
in the photospheric magnetogram
and arch-filament system in H$\alpha$
\citep{zwa85}.
\citet{str96} and \citet{str99} 
find the hierarchy of
motions of magnetic elements
in the active region:
faculae of positive and negative polarities
separate from each other
toward the edges of the region,
while pores of each polarity
move along the edges
toward the major sunspots.

Dynamics of the flux emergence
have been studied widely
through extensive
numerical simulations,
both in two and three dimensions
\citep[e.g.][]{shi89,fan01,arc04}.
It has been shown
that the flux tube rises
due to the Parker instability \citep{par66}
and expands into the corona
in a self-similar way,
which explains the characteristics
of the flux emergence
such as $\Omega$-shaped loops
and the downflows at the footpoints of the loops.

In this paper,
we report
first results of
the large-scale three-dimensional
magnetohydrodynamic (3D MHD) simulation
on the emergence of a twisted flux tube
from a deeper convection zone ($-20,000\ {\rm km}$)
into the corona through the photosphere.
In our previous 2D experiments,
\citet{tor10} and \citet{tor11b}
(hereafter papers I and I\hspace{-.1em}I, respectively),
we found the conditions
of the magnetic flux
for reasonable emergence
such as the field strength, the total flux,
and the twist of the tube
at a depth of $-20,000\ {\rm km}$.
We apply these values
as the initial conditions
in the 3D calculation.
As a result,
several characteristics are
found to be consistent
with the previous 2D experiments
and observations.
On the basis of this simulation,
we also suggest
a new theoretical picture
of the flux emergence
and the birth of
an active region
through the surface.

\section{Numerical Setup\label{sec:setup}}

The basic MHD equations
solved in this simulation
in vector form are as follows:
\begin{eqnarray}
  \frac{\partial\rho}{\partial t}
  + \mbox{\boldmath $\nabla$}
    \cdot(\rho\mbox{\boldmath $V$})=0,
\end{eqnarray}
\begin{eqnarray}
  \frac{\partial}{\partial t}
  (\rho\mbox{\boldmath $V$})
  + \mbox{\boldmath $\nabla$}\cdot
  \left(
    \rho\mbox{\boldmath $V$}\mbox{\boldmath $V$}
    +p\mbox{\boldmath $I$}
    -\frac{\mbox{\boldmath $BB$}}{4\pi}
    +\frac{\mbox{\boldmath $B$}^{2}}{8\pi}\mbox{\boldmath $I$}
  \right)
  -\rho\mbox{\boldmath $g$}=0,
\end{eqnarray}
\begin{eqnarray}
  \frac{\partial\mbox{\boldmath $B$}}
       {\partial t}
  = \mbox{\boldmath $\nabla$}
    \times (\mbox{\boldmath $V$}\times\mbox{\boldmath $B$}),
\end{eqnarray}
\begin{eqnarray}
  &&\frac{\partial}{\partial t}
  \left(
    \rho U
    + \frac{1}{2}\rho\mbox{\boldmath $V$}^{2}
    + \frac{\mbox{\boldmath $B$}^{2}}{8\pi}
  \right) \nonumber \\
  &&+\mbox{\boldmath $\nabla$}\cdot
  \left[
    \left(
      \rho U + p + \frac{1}{2}\rho\mbox{\boldmath $V$}^{2}
    \right)
    \mbox{\boldmath $V$}
    + \frac{c}{4\pi}
      \mbox{\boldmath $E$}\times\mbox{\boldmath $B$}
  \right]
  -\rho\mbox{\boldmath $g$}\cdot\mbox{\boldmath $V$}
  =0,
\end{eqnarray}
and
\begin{eqnarray}
  U=\frac{1}{\gamma-1}
    \frac{p}{\rho},
\end{eqnarray}
\begin{eqnarray}
  \mbox{\boldmath $E$}
  =-\frac{1}{c}
   \mbox{\boldmath $V$}\times\mbox{\boldmath $B$},
\end{eqnarray}
\begin{eqnarray}
  p=\frac{k_{\rm B}}{m}\rho T,
\end{eqnarray}
where $U$ is the internal energy per unit mass,
$\mbox{\boldmath $I$}$
the unit tensor,
$k_{\rm B}$ the Boltzmann constant,
$m$ the mean molecular mass,
and $\mbox{\boldmath $g$}$
the uniform
gravitational acceleration.
Other symbols have their usual meanings:
$\rho$ is for density,
$\mbox{\boldmath $V$}$ velocity vector,
$p$ pressure,
$\mbox{\boldmath $B$}$ magnetic field,
$c$ speed of light,
$\mbox{\boldmath $E$}$ electric field,
and $T$ temperature.
Here, the medium is assumed to be
an inviscid perfect gas
with a specific heat ratio $\gamma =5/3$.
The equations are normalized
by the pressure scale height
$H_{0}=200\ {\rm km}$ for length,
sound speed $C_{\rm s0}=8\ {\rm km\ s}^{-1}$ for velocity,
$\tau_{0}\equiv H_{0}/C_{\rm s0}=25\ {\rm s}$ for time,
$\rho_{0}=1.4\times 10^{-7}\ {\rm g\ cm}^{-3}$ for density,
all of which 
are typical values for the photosphere.
The units for pressure, temperature,
and magnetic field strength are
$p_{0}=9.0\times 10^{4}\ {\rm dyn\ cm}^{-2}$,
$T_{0}=4000\ {\rm K}$,
and $B_{0}=300\ {\rm G}$, respectively.

In this experiment,
we set 3D Cartesian coordinates $(x,y,z)$,
where $z$ is parallel
to the gravitational acceleration,
i.e., $\mbox{\boldmath $g$}=(0,0,-g_{0})$.
The numerical domain is
$(-400,-200,-200)\leq (x/H_{0},y/H_{0},z/H_{0})\leq (400,200,250)$
and the total grid number is
$1602\times 256\times 1024$.
The grid spacing
for $x$-direction is
$\Delta x/H_{0}=0.5$
(uniform);
for the other directions,
$\Delta y/H_{0}=0.5$
and $\Delta z/H_{0}=0.2$
at the central area of the domain,
which gradually increase
for each direction.
We assume periodic boundaries
for both horizontal directions
and symmetric boundaries
for the vertical direction.

\begin{figure}
  \resizebox{\hsize}{!}{\includegraphics[clip]{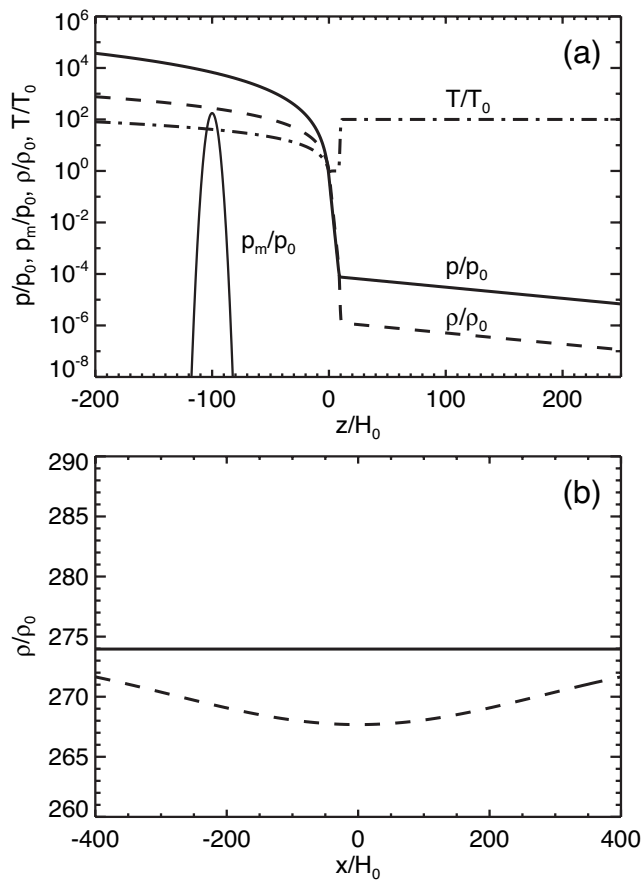}}
  \caption{(a) One-dimensional
    ($z$-)distributions
    of the initial background gas pressure
    (thick solid line),
    density (dashed line),
    and temperature (dash-dotted line).
    The magnetic pressure
    $p_{\rm m}=B^{2}/(8\pi)$
    along the vertical axis
    $x/H_{0}=y/H_{0}=0$
    is overplotted
    with a thin solid line.
    (b) Background density profile
    at $z=-100H_{0}$ (solid line).
    Density inside the tube
    along $(y,z)=(0,-100H_{0})$
    is overplotted
    with dashed line.}
  \label{fig:initial}
\end{figure}

The background stratification
consists of the adiabatically stratified
convective layer ($z/H_{0}<0$),
the cool isothermal ($T(z)/T_{0}=1$)
photosphere/chromosphere ($0\leq z/H_{0}< 10$),
and the hot isothermal ($T(z)/T_{0}=100$)
corona ($z/H_{0}> 10$).
Here, the photosphere is isothermally stratified,
i.e., convectively stable
against the rising motion of the plasma.
A transition region connects
the low- and high-temperature atmospheres
with a steep temperature gradient
at around $z/H_{0}=10$.
The initial flux tube
embedded at $z=-100H_{0}=-20,000\ {\rm km}$
is uniformly twisted,
i.e., $B_{\phi}(r)=qrB_{x}(r)$,
where $B_{x}(r)=B_{\rm tube}\exp{(-r^{2}/R_{\rm tube}^{2})}$
is the axial field,
$r$ the radial distance from the axis,
$B_{\rm tube}=67B_{0}=2.0\times 10^{4}\ {\rm G}$
the field strength at the axis,
$R_{\rm tube}=5H_{0}=1000\ {\rm km}$
the typical tube radius,
$B_{\phi}(r)$ the azimuthal component,
and $q$ the twist parameter
which is set to be
$0.1/H_{0}=5.0\times 10^{-4}\ {\rm km}^{-1}$
(stable to the kink instability).
The total flux of the axial component
is $\Phi=6.3\times 10^{20}\ {\rm Mx}$.
The middle of the tube
around $x/H_{0}=0$ is made buoyant
through thermal equilibrium
between the inside and
the outside of the tube,
and the buoyancy decreases
as a function of the form
$\exp(-x^{2}/\lambda^{2})$,
where $\lambda/H_{0}=400$.
The tube strength, the total flux, and the twist
are chosen to satisfy the conditions
found in papers I and I\hspace{-.1em}I.
The initial background distribution
of gas pressure, density, and temperature,
and the magnetic pressure
are indicated
in Figure \ref{fig:initial}(a).
The density
inside and outside
the tube
along $x$-axis
are shown
in Figure \ref{fig:initial}(b).
One can see that,
although the density is reduced
throughout the whole tube,
the middle part is
most buoyant.

\section{Results\label{sec:results}}

\subsection{General Evolution\label{sec:general}}

\begin{figure}
  \resizebox{\hsize}{!}{\includegraphics[clip]{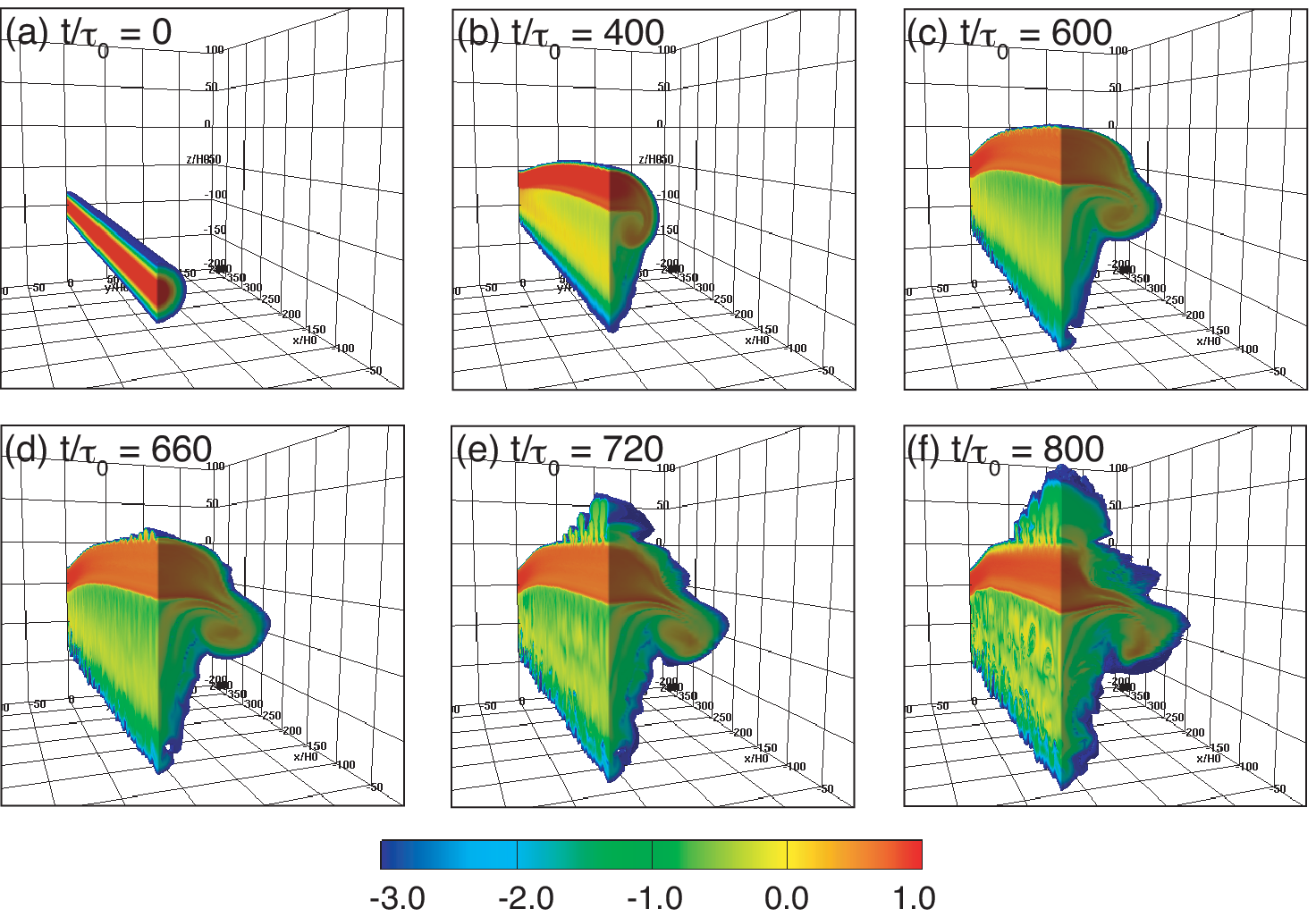}}
  \caption{Time-evolution of
      the twisted emerging flux tube.
      Logarithmic field strength
      $\log_{10}(|B|/B_{0})$
      for six different times
      in a limited region
      ($x/H_{0}<0, y/H_{0}>0$)
      is shown.
      The level of $z/H_{0}=0$
      corresponds to the boundary
      between the convection zone
      and the photosphere.}
  \label{fig:general}
\end{figure}

\begin{figure}
  \resizebox{\hsize}{!}{\includegraphics[clip]{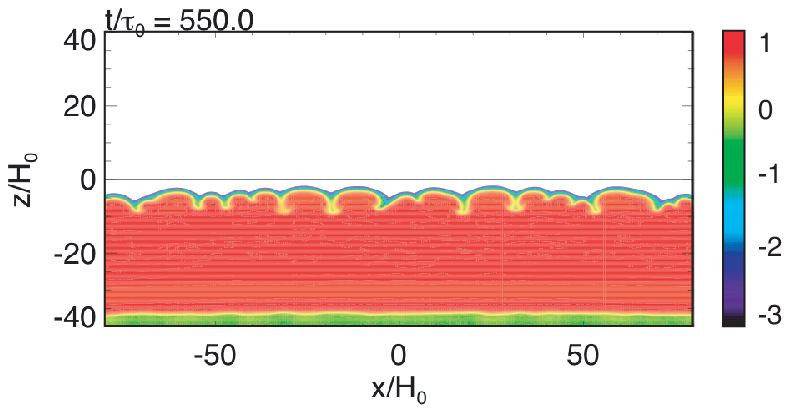}}
  \caption{Two-dimensional
    ($x$--$z$) closeup
    of the rising flux tube
    at $t/\tau_{0}=550$.
    Logarithmic field strength
    $\log_{10}(|B|/B_{0})$ is plotted,
    while the photospheric height
    $z/H_{0}=0$ is indicated
    with a solid line.}
  \label{fig:zoom}
\end{figure}

Figure \ref{fig:general} shows
the time evolution
of the emerging twisted flux tube.
Here, the logarithmic field strength
$\log_{10}{(|B|/B_{0})}$
only in the region $x/H_{0}\leq 0$
and $y/H_{0}\geq 0$ is plotted.
At $t/\tau_{0}=0$ (Fig. \ref{fig:general}(a)),
the initial tube is placed at $z/H_{0}=-100$.
Due to the buoyancy of the tube itself,
it rises through the convection zone
($t/\tau_{0}=400$,
see Fig. \ref{fig:general}(b)).
The tube expands
as the external density decreases
with height.
While the plasma draining
from the apex to both feet
accelerates
the rising tube,
the aerodynamic drag
decelerates the tube,
since the external flow
around the cross-section
forms a wake behind the main tube.
One can see a vortex roll
at the flank of the tube
and an elongated tail below.
However, the azimuthal component
of the flux tube yields
the inward curvature force
to maintain the tube coherent.
The upper surface
of the rising tube
becomes fluted
due to the interchange-mode
of the magnetic buoyancy instability
(magnetic Rayleigh-Taylor instability).
Figure \ref{fig:zoom}
is a two-dimensional ($x$--$z$) slice
of the rising flux tube,
which shows the fluting
at the top of the tube.
The rise velocity levels off
while in the middle
of the convection zone.

Approaching the surface at $t/\tau_{0}=600$
in Fig. \ref{fig:general}(c),
the tube decelerates
and expands laterally
in the $y$-direction
to make a flat structure
just beneath the photosphere
($60\ {\rm Mm}\times 20\ {\rm Mm}$).
The deceleration and the flattening occur
because the plasma
between the rising tube
and the convectively stable photosphere
is compressed,
which in turn suppresses
the rising tube from below.
It should be noted that
the deceleration and the flat magnetic structure
are consistent with the previous 2D experiments
(papers I and I\hspace{-.1em}I).
Here, the outermost field lines of the flat tube,
i.e., fields at the surface are
mostly in the negative $y$-direction.

As the surficial field strength increases
to satisfy the criterion
for the interchange-mode
of magnetic buoyancy instability
\citep[e.g.][]{new61},
the secondary evolution to the upper atmosphere
takes place at around $t/\tau_{0}=660$
(Fig. \ref{fig:general}(d)).
We find that
several
magnetic domes have been built
in the central area around $x/H_{0}=y/H_{0}=0$
at this stage,
aligned in the $x$-direction,
each being directed in the $y$-direction.
As the
central domes develop,
another
several
domes are newly created
beside the central
ones
($t/\tau_{0}=750$;
Fig. \ref{fig:general}(e)).
The domes continue growing
and merge with each other
in the corona.
After $t/\tau_{0}=700$,
the rising velocity declines again,
and eventually the emerging flux
reaches $z/H_{0}\sim 35$ 
at $t/\tau_{0}=800$
(Fig. \ref{fig:general}(f)).
The subsurface structure extends
$-200<x/H_{0}<200$ and $-40<y/H_{0}<40$.

The whole evolution process
is the same as the ``two-step emergence'' model
observed in the 2D experiments
(papers I and I\hspace{-.1em}I).
However, the final height of the coronal structure
is different from the 2D cases:
the height in the present calculation
is only $z/H_{0}=35$,
while the fluxes in 2D were more than 200.
The difference
occurs due to the three-dimensionality.
That is, in the 2D simulations,
the magnetic pressure
expands the magnetic flux
only in the $(x,z)$ or $(y,z)$ plane.
In the 3D case, however,
the magnetic pressure inflates the flux
in all $(x,y,z)$ directions,
which results in lower magnetic domes.
This 3D situation is the same as
the free expansion regime
by \citet{mat93}.

\subsection{Magnetic Structures in the Photosphere
  \label{sec:surface}}

\begin{figure}
  \resizebox{\hsize}{!}{\includegraphics[clip]{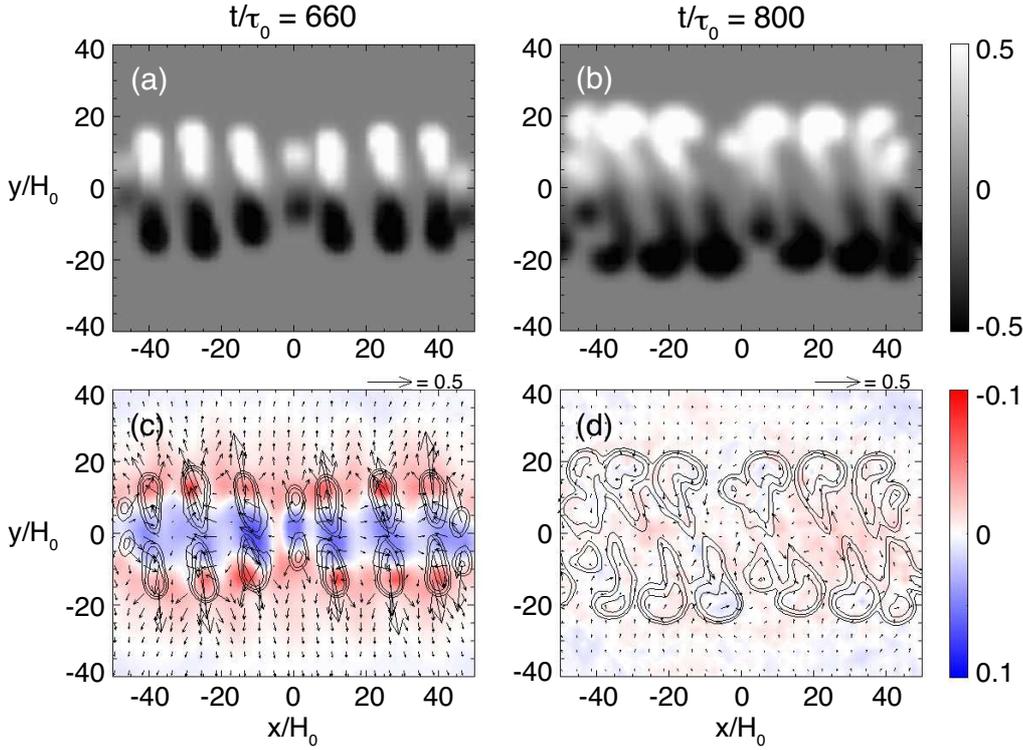}}
  \caption{({\it Top}) Time-evolution
      of vertical magnetic field strength
      $B_{z}/B_{0}$
      at the surface $z/H_{0}=0$
      (magnetogram).
      White (black) indicates
      the positive (negative) polarity.
      ({\it Bottom}) Corresponding velocity fields
      (Dopplergram).
      The vertical velocity
      $V_{z}/C_{\rm s0}$
      (color; red is downward)
      and the horizontal velocity
      $V_{\rm h}/C_{\rm s0}$
      (arrows)
      are shown.
      Corresponding vertical field
      $B_{z}/B_{0}$
      at the photosphere
      is over-plotted with contours.}
  \label{fig:surface}
\end{figure}

Figure \ref{fig:surface} shows
the vertical field strength
($B_{z}/B_{0}$; magnetogram)
and the corresponding vertical velocity
($V_{z}/C_{\rm s0}$; Dopplergram)
with the horizontal velocity field
($V_{\rm h}/C_{\rm s0}$)
at the surface $z/H_{0}=0$
at $t/\tau_{0}=660$ and 800.
In the Dopplergram,
red indicates a downward motion
($V_{z}/C_{\rm s0}<0$).
At $t/\tau_{0}\sim 610$,
we observe that
some blueshifts
and divergent flows appear
before the flux emergence,
which indicates that
the compressed plasma
between the rising tube and the photosphere
escapes laterally at the surface.

At $t/\tau_{0}=660$,
in Fig. \ref{fig:surface}(a),
magnetic elements
of positive and negative polarities
emerge onto the surface.
The absolute field strength
of each polarity
is more than a hundred Gauss.
In Fig. \ref{fig:surface}(c),
one can find
the blueshifts of the order
of a few ${\rm km\ s}^{-1}$
between each pair
and the redshifts
up to
$0.12C_{\rm s0}=1\ {\rm km\ s}^{-1}$
in the core of each patch.
At this time,
the horizontal speed
in the positive and negative polarities
is at its peak ($4$--$8\ {\rm km\ s}^{-1}$),
showing separative motions.
These features indicate that
the magnetic flux emerges upward,
while the plasma drains downward
to both footpoints
along the field lines.
Here, the surface field
is mostly directed
in the negative $y$-direction,
and the wavelengths are
$\lambda_{\parallel}\sim 20H_{0}$
and $\lambda_{\perp}\sim 15H_{0}$
respectively,
where $\lambda_{\parallel}$
and $\lambda_{\perp}$
are the wavelengths
parallel and perpendicular
to the surface horizontal field.
The parallel wavelength
$\lambda_{\parallel}\sim 20H_{0}$
is the most unstable wavelength
of the linear Parker instability
at the photosphere.

We speculate that
the wavelength
perpendicular to the field
$\lambda_{\perp}$
is determined
by the wavelength
of the interchange-mode instability
of the flux tube
before it reaches
the surface.
During its ascent
within the convection zone,
the surface of the flux tube
is found to be fluted
due to the interchange instability
(see Sect. \ref{sec:general}).
When the density smoothly increases upward
between the magnetized
and the unmagnetized atmosphere,
the growth rate
of the interchange instability
levels off as the wavenumber increases,
and the typical wavenumber
of this saturation
is approximately
an inverse
of the density transition scale
\citep{cha61}.
Considering the original tube
is assumed to have
a Gaussian profile,
i.e., $B_{x}(r)=B_{\rm tube}\exp{(-r^{2}/R_{\rm tube}^{2})}$,
and thus the density transition is
also a function of $\exp{(-r^{2}/R_{\rm tube}^{2})}$,
the scale of this transition layer is
$\sim R_{\rm tube}$.
Therefore,
the wavelength of this instability
within the convection zone
is approximately several times
the original tube's radius
($R_{\rm tube}=5H_{0}$),
which results in the wavelength
of the secondary emergence
$\lambda_{\perp}\sim 15H_{0}$
at the photosphere.

At $t/\tau_{0}=800$,
the magnetic pairs develop
and the region extends
$-120<x/H_{0}<120$ and $-30<y/H_{0}<30$,
i.e., $48\ {\rm Mm}\times 12\ {\rm Mm}$
(Fig. \ref{fig:surface}(b)).
The direction of separations
is found to be tilted.
Here, each separated patch
has formed a tadpole-like shape;
$|B_{z}|$ up to 350 G.
This configuration is formed
because the newly emerged elements
separate outward
and catch up with the elements
that emerged earlier,
and stop at the edge of the region.
Also,
the heads of
these tadpoles
make two alignments
at the edges,
and they show shearing motion.
The shearing
is leftward where $y/H_{0}>0$,
and rightward where $y/H_{0}<0$.
In this phase,
the total unsigned flux
$\int_{z=0} |B_{z}|\, dxdy$
reaches up to
$\sim 3.3\times 10^{20}\ {\rm Mx}$.
In Fig. \ref{fig:surface}(d)
the redshifts are
no longer seen,
which indicates that
the emergence has stopped.
To clarify the difference
between the aligned large elements
(tadpole-heads)
and the tilted elements
(tadpole-tails),
we use the term ``pores''
for the heads
of the tadpole-like features.
Moreover,
this term is used
in the sense
of the accumulated fields
that do not reach
the size of a sunspot.

\begin{figure}
  \resizebox{\hsize}{!}{\includegraphics[clip]{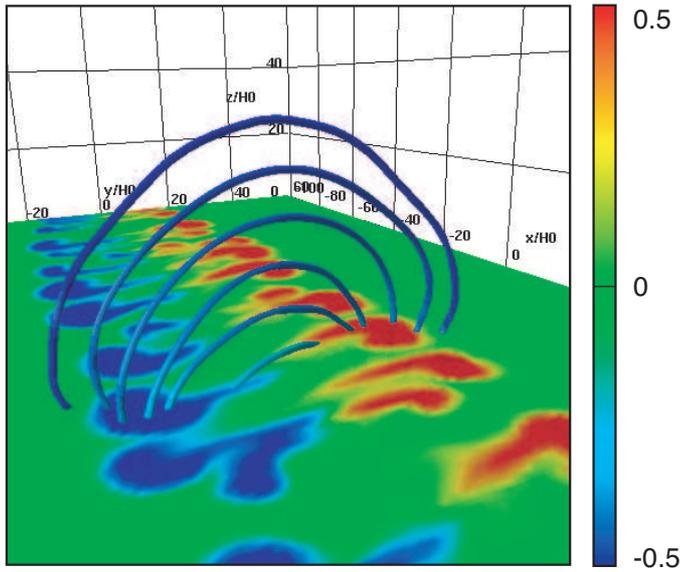}}
  \caption{Perspective view
      of selected field lines
      that pass through
      the vertical axis
      $(x/H_{0},y/H_{0})=(26,0)$
      at $t/\tau_{0}=800$,
      plotted on the photospheric vertical field
      $B_{z}/B_{0}$.}
  \label{fig:bvct}
\end{figure}

\begin{figure}
  \resizebox{\hsize}{!}{\includegraphics[clip]{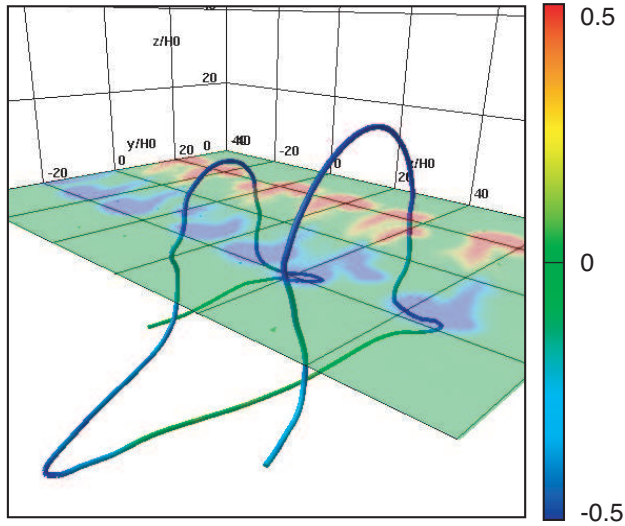}}
  \caption{A single field
    line that connects
    different photospheric magnetic elements.
    Plotted is the photospheric vertical field
    $B_{z}/B_{0}$
    at $t/\tau_{0}=800$,
    which is transparentized.}
  \label{fig:bvct_trans}
\end{figure}

Figure \ref{fig:bvct} shows
the selected field lines
plotted on the surficial magnetogram
at $t/\tau_{0}=800$.
Here,
emerged field lines in the corona
connect positive and negative polarities
in the photosphere.
The footpoints of coronal fields
stop 
at the edge of the region,
which is determined
by the extension
of the flat magnetic structure
beneath the surface.
That is,
the size of the active region
($48\ {\rm Mm}\times 12\ {\rm Mm}$
in this case)
depends on
the subphotospheric flux tube
($60\ {\rm Mm}\times 20\ {\rm Mm}$).
The wavelength of the 
initial density deficit,
$\lambda$,
would be one of the parameters
that determine the size of
the subphotospheric structure.
We also found that some field lines
connect different patches
deep under the surface
(see Figure \ref{fig:bvct_trans}).
Although such field lines are
not undulating
at around the surface,
they are still reminiscent
of a sea-serpent configuration
and a corresponding
resistive emergence process.

The tilt of magnetic elements
in the central area
is caused by
the emergence of the inner field lines
(kinematic effect).
Here, the initial flux tube
is uniformly twisted,
and thus the pitch angle of inner fields
are smaller.
Therefore,
the footpoints shift
as the inner field lines rise.
Also, the shearing of two aligned ``pores''
is due to
the Lorentz force
acting on the surface field
(dynamic effect;
see \citet{man01,fan01}).

\section{Summary and Discussion
\label{sec:discussion}}

In this paper,
we performed a 3D MHD simulation
of the emergence
of a twisted flux tube.
The initial tube
at $-20,000\ {\rm km}$
in the convection zone
has a field strength of
$2.0\times 10^{4}\ {\rm G}$,
a total flux of
$6.3\times 10^{20}\ {\rm Mx}$,
and a twist of
$5.0\times 10^{-4}\ {\rm km}^{-1}$,
which starts rising
due to its own magnetic buoyancy.
On reaching
the surface
after $t\sim 2.8\ {\rm hr}$,
the tube decelerates
and extends horizontally
($60\ {\rm Mm}\times 20\ {\rm Mm}$)
owing to the convectively stable photosphere
ahead of the tube.
As the surface field
satisfies the criterion
for the magnetic buoyancy instability,
the field emerges again
into the upper atmosphere
after $t\sim 4.1\ {\rm hr}$.
Eventually,
several magnetic domes
attain a height of
$\sim 7000\ {\rm km}$
at $t\sim 5.6\ {\rm hr}$.
The size of the active region
grows to $48\ {\rm Mm}\times 12\ {\rm Mm}$,
and the photospheric flux
amounts to
$3.3\times 10^{20}\ {\rm Mx}$.
This two-step feature
is consistent
with our 2D models
(papers I and I\hspace{-.1em}I).

We also observed
multiple
separation events
of the magnetic elements
at the photosphere,
posterior to the horizontal escaping flow
of the compressed plasma.
Such separating elements
move apart from each other
at the rate of $4$--$8\ {\rm km\ s}^{-1}$
and stop at the edges
of the active region,
which is determined
by the extension
of the subphotospheric field.
The multiple separations
are the results
of the interchange-mode instability
of the rising tube
while in the convection zone.
The magnetic elements then
gather at the edges
to make two alignments
of the ``pores''
(tadpole-heads).
The alignments
show shearing motions
at the rate of $\sim 0.5\ {\rm km\ s}^{-1}$,
which is explained
by the inner field emergence
(kinematic effect)
and the Lorentz force effect
(dynamic effect).
Upflows of a few ${\rm km\ s}^{-1}$
and downflows up to $1\ {\rm km\ s}^{-1}$
are observed in the emergent areas
and in the cores of surface 
fields respectively.
As far as we know,
we have never
observed
such photospheric features,
especially the
multiple separations,
in previous
3D calculations
applying flux tube as an initial condition
\citep[e.g.][]{fan01,arc04}.
Note that some calculations
using flux sheets showed
multiple separations
via similar instabilities
\citep{iso05,arc09}.

These features are strongly reminiscent
of the observations of NOAA AR 5617
by \citet{str96} and \citet{str99}.
They found that faculae of both polarities
separate from each other
toward the edges of the region,
which is similar to our findings
of multiple separations.
Also the shearing motions
of the aligned 
tadpole-heads
are consistent with
the pores moving along the edges
of the region toward
the main sunspots.
The difference of the tilt
of magnetic elements
between these two cases
is caused by the twist direction.
We assumed a right-handed tube
in the initial state,
which is favorable
for the southern hemisphere.
AR 5617 appeared
in the northern hemisphere,
which yields the left-handed twist.

The summary of the comparison
between our results
and the observations
by \citet{str96} and \citet{str99}
is presented in Table \ref{tab:comparison}.
The size of AR
and the velocities
are consistent 
between the two.
As for the difference
of separation speeds,
note that \citet{har73} observed
that faculae of opposite polarities
separate initially at the rate of
$> 2\ {\rm km\ s}^{-1}$.
The photospheric total flux is
one digit smaller
than the observed value,
and we did not find
any major sunspots
in our active region.
These differences
may be because,
in our calculation,
the rising tube
stops in 1.4 hours
after it appears at the surface.
The age of AR 5617
was estimated to be 6.5--7.5 hours old
at the beginning of the observation
and 8--9 hours old at the end.
In the observation,
emergence events showed
undulatory structures
with a typical wavelength
of $8\ {\rm Mm}$,
while, in our case,
each field line
does not show undulation
at around the surface.
Such undulating features
might be found
in the more resolved calculations
or in the calculations
including thermal convection
\citep{iso07,che10}.
\citet{str99} summarized
their observations
as a model
in which
each emergence event (separation)
occurs in a single vertical sheet,
which
forms a series
of sheets aligned in a parallel fashion
(see Fig. 8 of their paper).
From our results,
the separations are explained as
the consequence
of the interchange-mode instability
of the flattened flux tube
beneath the surface
(see e.g. Fig. \ref{fig:zoom}).

\begin{table}
  \caption{Summary of Comparison
    with AR 5617\label{tab:comparison}}
  \centering
  \begin{tabular}{lcc}
    \hline\hline
    & Simulation Results & AR 5617\tablefootmark{a}\\
    \hline
    size of the region & $48\ {\rm Mm}\times 12\ {\rm Mm}$ &
    $50\ {\rm Mm}\times 30\ {\rm Mm}$\\
    vertical unsigned total flux &
    $3.3\times 10^{20}\ {\rm Mx}$ &
    $4\times 10^{21}\ {\rm Mx}$\\
    age from the appearance &
    $0$--$1.4\ {\rm hr}$ & $6.5$--$9\ {\rm hr}$ \\
    separation speed & $4$--$8\ {\rm km\ s}^{-1}$ &
    $0.84\ {\rm km\ s}^{-1}$
    \tablefootmark{b} \\
    shearing speed & $\sim 0.5\ {\rm km\ s}^{-1}$ &
    $0.73\ {\rm km\ s}^{-1}$
    \tablefootmark{c} \\
    upflow velocity & ${\rm a\ few\ km\ s}^{-1}$ &
    $0.86\ {\rm km\ s}^{-1}$
    \tablefootmark{d} \\
    downflow velocity & $\lesssim 1\ {\rm km\ s}^{-1}$ &
    $1.26\ {\rm km\ s}^{-1}$
    \tablefootmark{e} \\
    wavelength of emergence pattern & - &
    $\sim 8\ {\rm Mm}$ \\
    \hline
  \end{tabular}
  \tablefoot{
    \tablefoottext{a}{\citet{str96} and \citet{str99}.}
    \tablefoottext{b}{Separation of facular elements.}
    \tablefoottext{c}{Separation of pores.}
    \tablefoottext{d}{Upflow in an emergent region.}
    \tablefoottext{e}{Downflow in a facula.}
    }
\end{table}

\begin{figure}
  \resizebox{\hsize}{!}{\includegraphics[clip]{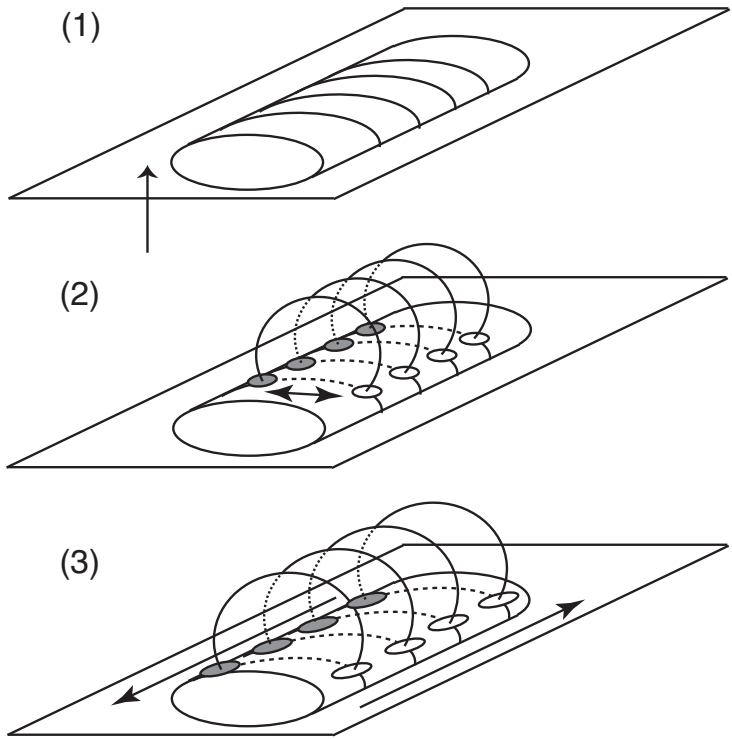}}
  \caption{Picture of flux tube emergence
      and the formation of
      an active region
      derived from the calculation results
      in this paper.
      Solar surface is indicated
      by a horizontal slice.
      (1) Flux tube rising through
      the convection zone
      decelerates
      to form a flat structure
      beneath the surface.
      (2) Magnetic elements
      of opposite polarities
      are observed to separate
      due to the interchange-mode instability.
      White and black ellipses indicate
      the positive and negative elements.
      These elements gather
      at the edges of the active region
      to make two alignments of pores.
      (3) Aligned pores
      show shearing motion
      when inner fields emerge.
    }
  \label{fig:illustration}
\end{figure}

On the basis
of the numerical results
in this paper,
we suggest a theoretical picture
of the flux emergence
and the formation of
an active region
through the surface,
which includes
\citet{str99}'s model.
Our model is schematically illustrated
in Fig. \ref{fig:illustration}.
(1) The flux tube rises
through the convection zone
due to magnetic buoyancy.
On approaching
the surface,
the tube decelerates
and becomes flattened
because of the photosphere
in front of the rising tube.
(2) When the photospheric field
satisfies the criterion
for the magnetic buoyancy instability,
further evolution to the corona
breaks out.
Due to the interchange-mode instability,
photospheric magnetogram shows
multiple separation events.
The separated elements
reach the edges of the region,
whose size is determined
by the subsurface field.
These elements then make
two alignments of pores.
Compare this panel
with Fig. 8
of \citet{str99}.
(3) As the emergence continues,
inner fields of the flux tube rise
and footpoints shift
to show shearing motions.
Lorentz force also drives
the pores
to shear.

The initial condition
of the present experiment
($B_{\rm tube}=67B_{0}=2.0\times 10^{4} G$,
$R_{\rm tube}=5H_{0}=1000\ {\rm km}$,
and $q=0.1/H_{0}=5.0\times 10^{-4}\ {\rm km}^{-1}$)
is the same as that of Case 5
of our 2D cross-sectional calculation
in \citet{tor11b}.
One of the basic differences
between these two simulation results
is the rising time
from the initial depth
of $-20\ {\rm Mm}$
to the surface.
In 2D calculation,
the flux tube
reached the surface
in $t\sim 550\tau_{0}=3.8\ {\rm hr}$,
while, in 3D case,
it took $t\sim 600\tau_{0}=4.2\ {\rm hr}$.
That is,
the 3D tube rises slower.
In 3D case,
plasma in the tube apex
drains down
along the field lines
to the both feet of the tube,
which drives the tube
more buoyant.
At the same time,
in the 3D regime,
the magnetic curvature force
pulls down the rising tube.
The time difference
between the two cases
indicates that
the curvature force
dominates the draining effect
in the present 3D experiment.

In the solar interior
as well as in the photosphere,
several classes of convection
may affect the rise of magnetic flux.
Recent observations
by SOT on board {\it Hinode} satellite
have revealed the convective nature
of the surface field
\citep{lit09}.
Stronger pores
($> 1 {\rm kG}$) or sunspots,
which are not found
in this calculation
without convection,
would be formed through the convective
collapse process
\citep{par78}.
Large-scale upflow supports
the rising of the tube,
and more flux could be transported
to the surface
\citep{fan03}.
Surface convection creates
undulating fields,
and the cancellation
of such fields
removes the mass
trapped in the U-loop
\citep{iso07,che10}.
This process accelerates
the density draining
from the surface layer,
which may be important to
the spot formation.
On the other hand,
\citet{ste11} reported that
flux of 20 kG
in their convective experiment,
the same as used in ours,
is too strong,
because it produces large,
hot, bright granules
at the surface,
which are not seen
in the Sun.
This is
an interesting difference
between the two types
of calculations.
It should be noted
that the initial settings are different
in the two cases;
we used
an initial horizontal flux tube
in the convectively stable interior,
while, in \citet{ste11}'s calculation,
uniform horizontal fields
are advected into the computational domain
by the convective upflows
from the bottom boundary.
More theoretical and observational
studies are needed
on the effects
of the convection
in the whole flux emergence process.

\begin{acknowledgements}
  Numerical computations
  were carried out on Cray XT4
  at the Center for Computational Astrophysics, CfCA,
  of the National Astronomical Observatory of Japan.
  The page charge
  of this paper
  is supported by CfCA.
  This work was supported
  by the JSPS Institutional Program
  for Young Researcher Overseas Visits,
  and by the Grant-in-Aid for JSPS Fellows.
  We thank Dr. Y. Fan
  of the High Altitude Observatory,
  the National Center
  for Atmospheric Research,
  and the anonymous referee
  for improving
  this paper.
  We are grateful
  to the GCOE program instructors
  of the University of Tokyo
  for proofreading/editing assistance.
\end{acknowledgements}

\bibliographystyle{aa}
\bibliography{reference}

\end{document}